\documentclass[a4paper,11pt]{article}

\usepackage{contribution}

\newcommand{\weblink}[2][]{%
    \ifthenelse{\equal{#1}{}}%
    {\textnormal{\url{#2}}}%
    {\textnormal{\href{#2}{#1}}}%
}

\def\beq{\begin{equation}}
\def\eeq#1{\label{#1}\end{equation}}
\def\eeqn{\end{equation}}

\def\beqa{\begin{eqnarray}}
\def\eeqa#1{\label{#1}\end{eqnarray}}
\def\eeqan{\end{eqnarray}}

\let\bar=\overbar

\def\Dslash{\not{\hbox{\kern-4pt $D$}}}
\def\dslash{\not{\hbox{\kern-2pt $\del$}}}

\def\msb{{\bar{\ssstyle M \kern -1pt S}}}

\newcommand{\contribution}[7][]{%
  \clearpage
  \thispagestyle{plain}
  \ifthenelse{\equal{#1}{}}
  {\hypersetup{pdftitle={#2}}}
  {\hypersetup{pdftitle={#1}}}
  \hypersetup{pdfauthor={{#3} {#4}}}
  {\centering\normalfont\LARGE\bfseries\sffamily #2 \par\nobreak}
  \lhead{}
  \chead{%
    \textit{\footnotesize XIV International Conference on Hadron Spectroscopy
      (\weblink[\textit{hadron2011}]{http://www.hadron2011.de}), 13-17 June 2011, Munich, Germany}%
  }
  \rhead{}
  \bigskip
  \begin{center}
    {#3} {#4}\ifthenelse{\equal{#6}{}}{}{\footnote{\weblink[#6]{mailto:#6}}}
    \ifthenelse{\equal{#7}{}}{}{#7} \\
    \textit{#5}
  \end{center}
  \bigskip
}

\renewcommand{\abstract}[1]{%
  \begin{center}
    \begin{minipage}{0.85\textwidth}
      \begin{footnotesize}
        #1
      \end{footnotesize}
    \end{minipage}
  \end{center}
  \bigskip
}

\begin{document}

%
%
%
%
%
{  


%

\makeatletter
\@ifundefined{c@affiliation}%
{\newcounter{affiliation}}{}%
\makeatother
\newcommand{\affiliation}[2][]{\setcounter{affiliation}{#2}%
  \ensuremath{{^{\alph{affiliation}}}\text{#1}}}
%

\contribution[Pion scattering and electro-production on nucleons]
{Pion scattering and electro-production on nucleons in the resonance
region in chiral quark models}
{Simon}{\v{S}irca}  
{\affiliation[Faculty of Mathematics and Physics, University of Ljubljana, Slovenia]{1} \\
 \affiliation[Jo\v{z}ef Stefan Institute, Ljubljana, Slovenia]{2} \\
 \affiliation[Faculty of Education, University of Ljubljana, Slovenia]{3} \\
 \affiliation[Departamento de Fisica and Centro de Fisica Computacional,
 Universidade de Coimbra, Portugal]{4}}
{simon.sirca@fmf.uni-lj.si}
{\!\!$^,\affiliation{1}\affiliation{2}$,
Bojan Golli\affiliation{3}\affiliation{2},
Manuel Fiolhais\affiliation{4},
and Pedro Alberto\affiliation{4}}

\abstract{%
Pion scattering and electro-production amplitudes have been
computed in a coupled-channel framework incorporating quasi-bound
quark-model states, based on the Cloudy Bag model.
All relevant low-lying nucleon resonances in the P33,
P11, and S11 partial waves have been covered, including
the $\Delta(1232)$, the $N^*(1440)$, $N^*(1535)$, and $N^*(1650)$.  
Consistent results have been obtained for elastic and inelastic
scattering (two-pion, eta-N, and K-Lambda channels),
as well as for electro-production.
The meson cloud has been shown to play a major role,
in particular in electro-magnetic observables in the P33 and P11 channels.
}
%


The study of pion scattering and electro-production on nucleons
in the region of low-lying resonances helps us establish meaningful
connections between the data and resonance properties obtained
in model calculations.  One of the key aspects
is the interplay of quark and meson degrees of freedom.
The conceptual foundations of our approach in chiral quark models
date back to the paper \cite{plb96}, in which we demonstrated
the importance of the pion cloud in the electro-production
of pions in the region of the $\Delta(1232)$.  We found that
the pion cloud contributes $\sim 45\,\%$ to the magnetic
dipole amplitude, and strongly dominates the electric quadrupole
amplitude, a result later confirmed by many related calculations.

In our previous work  \cite{epja08} we have shown that in models
in which mesons couple linearly to the quark core, the elements
of the $K$~matrix for meson-baryon scattering $MB\rightarrow M'B'$
in the basis with good total angular momentum $J$, isospin $T$
and parity take the form:
\begin{equation}
   K_{M'B'\,MB}^{JT} =  -\pi\mathcal{N}_{MB}
     \langle\Psi^{MB}_{JT}||V_{M'}(k)||{\Psi}_{B'}\rangle\,,
\qquad
   \mathcal{N}_{MB} = \sqrt{\omega_{M} E_{B} \over k_{M} W}\, .
\label{defK}
\end{equation}
Here $|{\Psi}^{MB}_{JT}\rangle$ is the principal value state:
$$
|\Psi^{MB}_{JT}\rangle = \mathcal{N}_{MB}\left\{
    [a^\dagger(k_M)|{\Psi}_B\rangle]^{JT} 
+
 \sum_{\mathcal{R}}c_{\mathcal{R}}^{MB}|\Phi_{\mathcal{R}}\rangle
+ \sum_{M'B'}
   \int {\mathrm{d} k\>
       \chi^{M'B'\,MB}(k)\over\omega_k+E_{B'}(k)-W}\,
      [a^\dagger(k)|{\Psi}_{B'}\rangle]^{JT}\right\} \>,
$$
$\Psi_{B'}$ is the fully dressed outgoing baryon state,
and $V_{M'}(k)$ is the quark-meson vertex determined
in the underlying quark model.
The first term represents the free meson ($\pi$, $\eta$, $\sigma$, ...)
and the baryon ($N$, $\Delta$, ...) and defines the channel,
the next term is the sum over {\em bare\/} tree-quark states 
$\Phi_{\mathcal{R}}$ involving different excitations of the quark 
core, the third term introduces meson clouds around different isobars.
In the case of two-pion decays, we assume that they proceed
either through an unstable meson  (e.g.~$\rho$-meson, $\sigma$-meson)
or through a baryon resonance (e.g.~$\Delta(1232)$, $N^*(1440)$).
The meson amplitudes $\chi^{M'B'\,MB}(k)$ are proportional to the
(half) off-shell $K$-matrix elements (\ref{defK}), and obey
a coupled set of equations of the Lippmann-Schwinger type.
The resulting amplitudes take the form
$$
   \chi^{M'B'\,MB}(k) 
     = -\sum_{\mathcal{R}}{c}^{MB}_{\mathcal{R}}
    {\cal V}^{M'}_{B'\mathcal{R}}(k)
       + \mathcal{D}^{M'B'\,MB}(k)\,,
$$
where the first term represents the contribution of
resonances, while $\mathcal{D}^{M'B'\,MB}(k)$ 
originates in non-resonant background processes.
Here $  {c}_{\mathcal{R}}^{MB} 
          = {{\cal V}^M_{B\mathcal{R}}
             / (Z_{\mathcal{R}}(W) (W-M_{\mathcal{R}})) }$,
${\mathcal{V}}^M_{B\mathcal{R}}$ is the dressed
matrix element of the quark-meson interaction between
the resonant  state and the baryon state in the channel $MB$, 
and $Z_{\mathcal{R}}$ is the wave-function normalization. 
The resulting physical resonant states $\mathcal{R}$ are
superpositions of the bare 3-quark states $\Phi_{\mathcal{R}'}$.
The $T$ matrix is obtained by solving the Heitler's equation
$T = K + \mathrm{i}TK$.

\begin{figure}[htb]
  \begin{center}
    \includegraphics[width=0.49\textwidth]{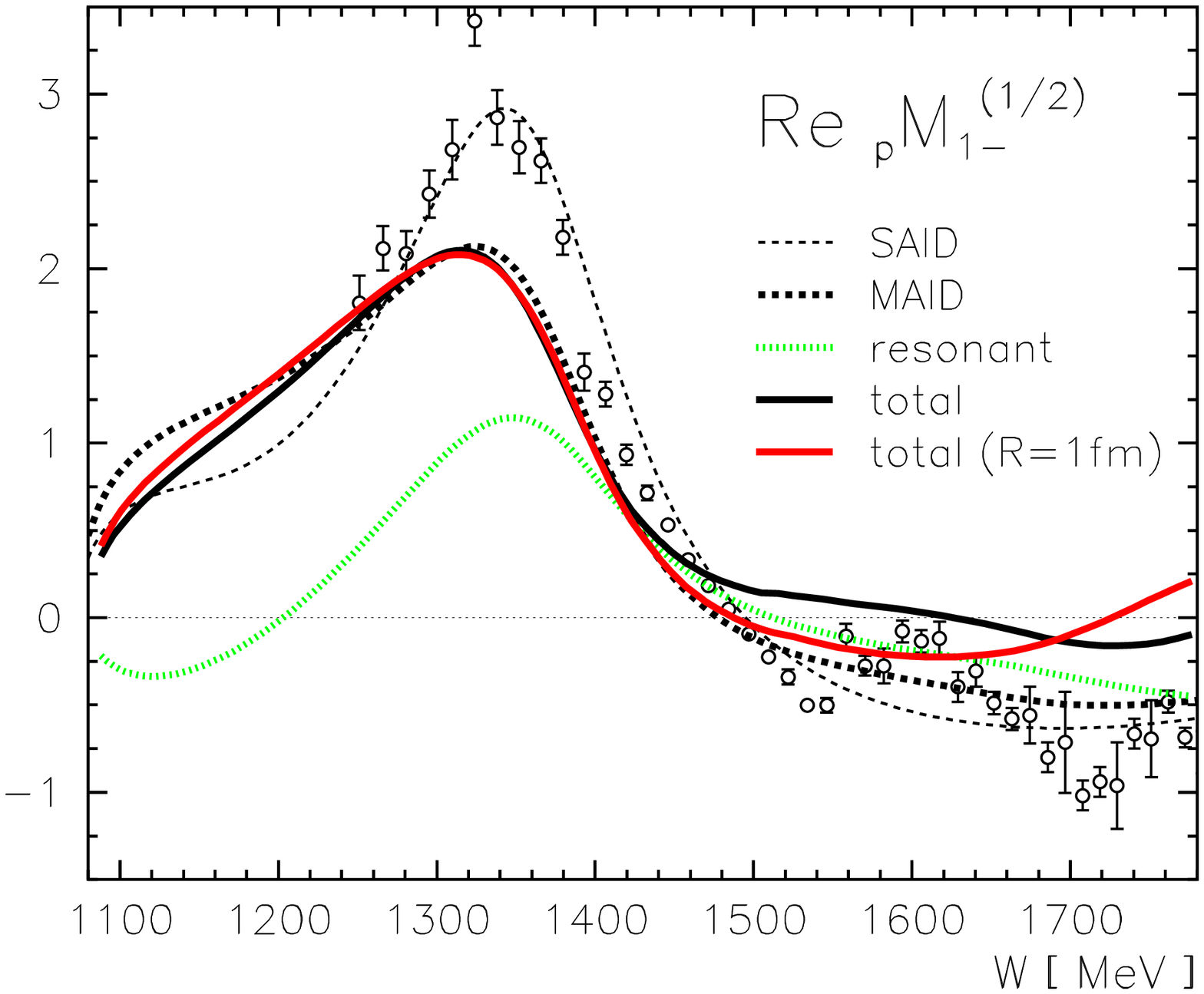}
    \includegraphics[width=0.49\textwidth]{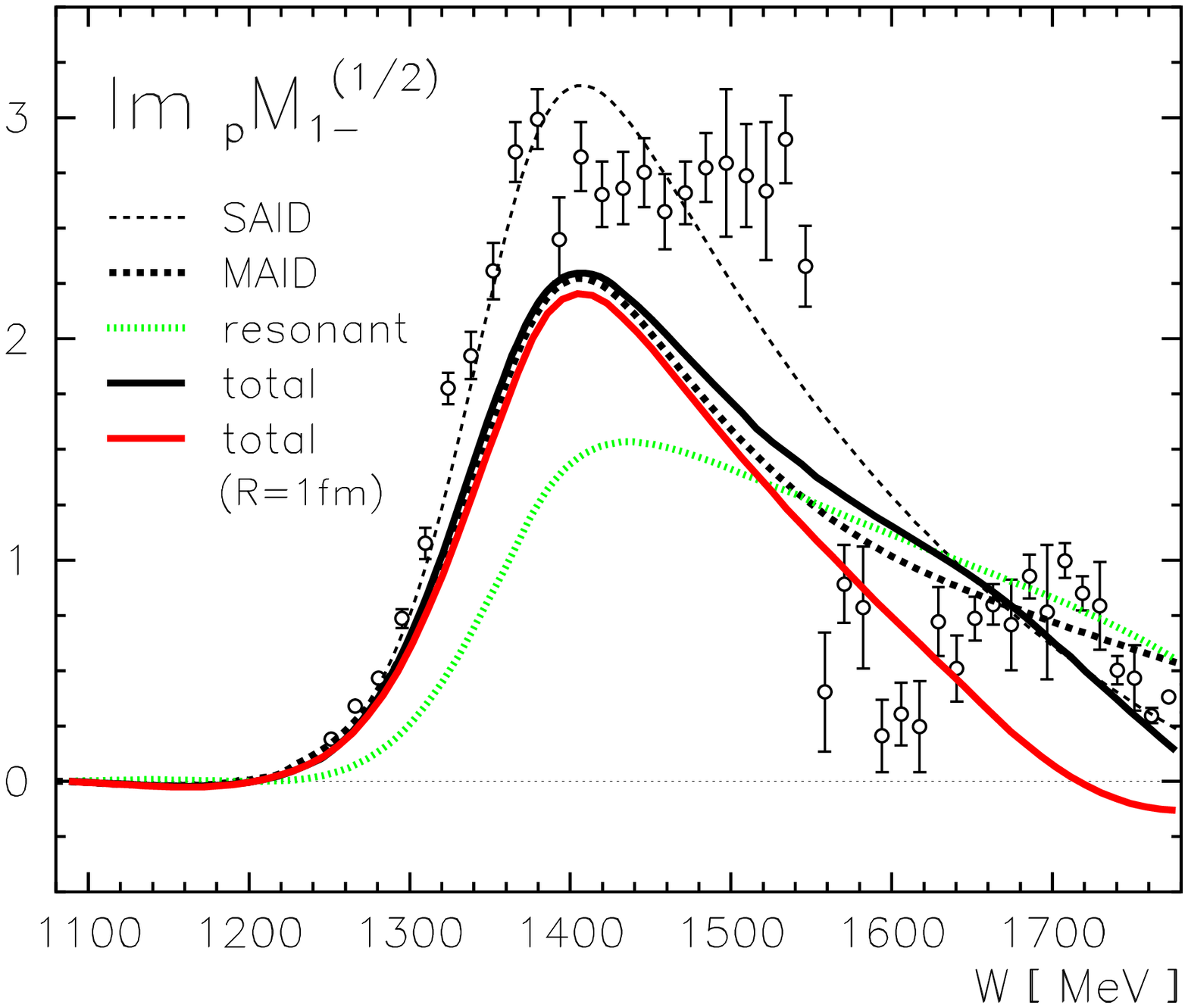}
    \vspace*{-3mm}
    \caption{Real and imaginary parts of the magnetic electro-production
multipole for the proton target, isospin-$1/2$ channel, in the P11
partial wave (the Roper resonance).}
    \label{fig:pM1m}
  \end{center}
\end{figure}

More recently, we have extended the method outlined above
to the calculation of meson electro-production amplitudes
by including the $\gamma N$ channel \cite{epja09,epja11}.
The electro-production amplitude in the vicinity of
a chosen resonance ${\cal R}=N^*$ can be cast in the form
$$
\mathcal{M}_{\gamma NMB}  = 
\mathcal{M}_{\gamma NMB}^{(res)} +
\mathcal{M}_{\gamma NMB}^{(bg)}  = 
\sqrt{\omega_\gamma E_N^\gamma \over \omega_M E_B }\,
{1\over\pi{\cal V}_{BN^*}}\,
   \langle\Psi_{N^*}^{(res)}(W)|{V}_\gamma
                 |\Psi_N\rangle\, T_{MBMB}
+ \mathcal{M}_{\gamma NMB}^{(bg)} \>, 
$$
where $T_{MBMB}$ is the meson-baryon scattering
amplitude, and the electro-excitation of the resonance is
described by the helicity amplitude
$\langle\Psi_{N^*}^{(res)}(W)|{V}_\gamma|\Psi_N\rangle 
= A_{\gamma N\to N^*}$.   
Here ${{V}_\gamma(\mu,\vec{k}_\gamma)}$
is the interaction of the photon with the electro-magnetic
current, and contains quark and pion contributions
as specified by the underlying quark model.

\begin{figure}[htb]
  \begin{center}
    \includegraphics[width=0.47\textwidth]{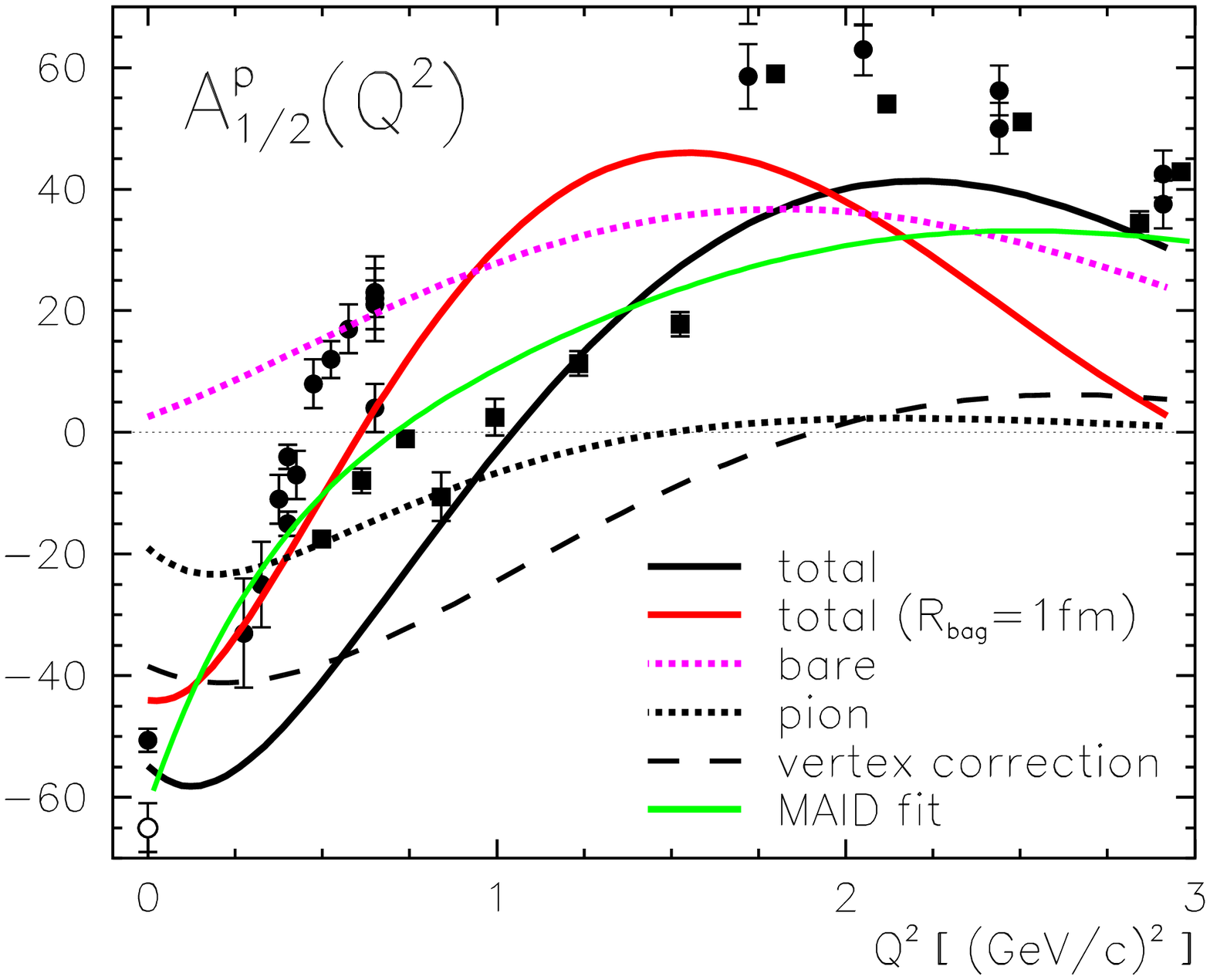}
    \includegraphics[width=0.47\textwidth]{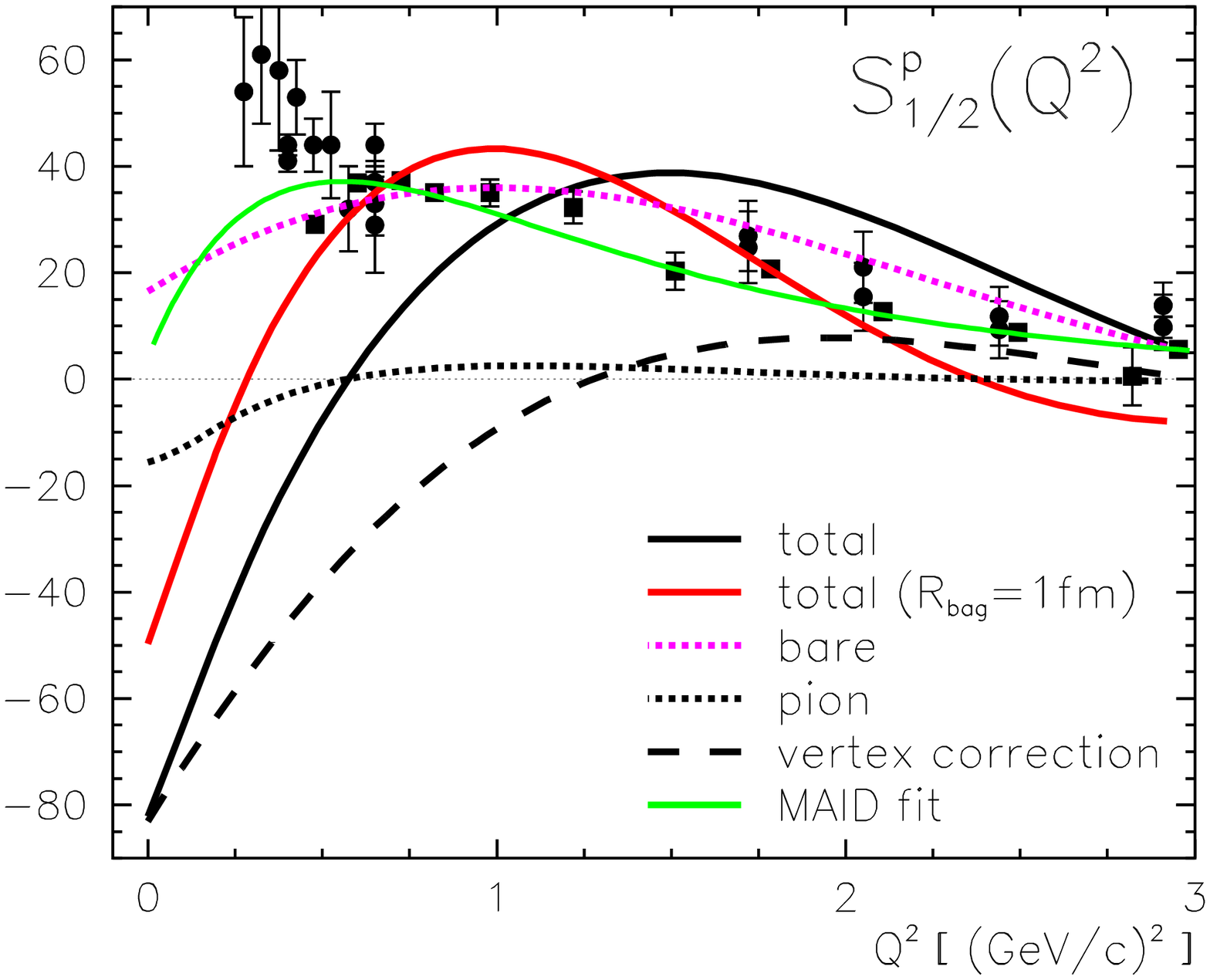}
    \vspace*{-3mm}
    \caption{Different contributions to the transverse and scalar
helicity amplitudes for the electro-excitation of the $N^\star(1440)$
(Roper) resonance in the P11 partial wave.}
    \label{fig:hel1520}
  \end{center}
\end{figure}

In the P11 case we have included the $\pi N$,
$\pi\Delta$, $\pi N^*(1440)$ and $\sigma N$ channels,
while the sum over $\mathcal{R}$ consists of the nucleon,
$N^*(1440)$, and $N^*(1710)$ resonance;
in the S11 case we have considered the $\pi N$,
$\pi\Delta$, $\pi N^*(1440)$, $\rho N$ and $K\Lambda$ channels,
as well as $N^*(1535)$ and $N^*(1650)$ resonances.
We have used the quark-model wave-functions for
the negative-parity states in the form
$
\Phi_{\cal R} 
  = c^{\cal R}_{A} \vert (1s)^2(1p_{3/2})^1 \rangle
    + c^{\cal R}_{P} \vert (1s)^2(1p_{1/2})^1 \rangle_1
    + c^{\cal R}_{P'} \vert (1s)^2(1p_{1/2})^1 \rangle_2$,
where the mixing coefficients of \cite{myhrer} have been adopted.

\begin{figure}[htbp]
  \begin{center}
    \includegraphics[width=0.47\textwidth]{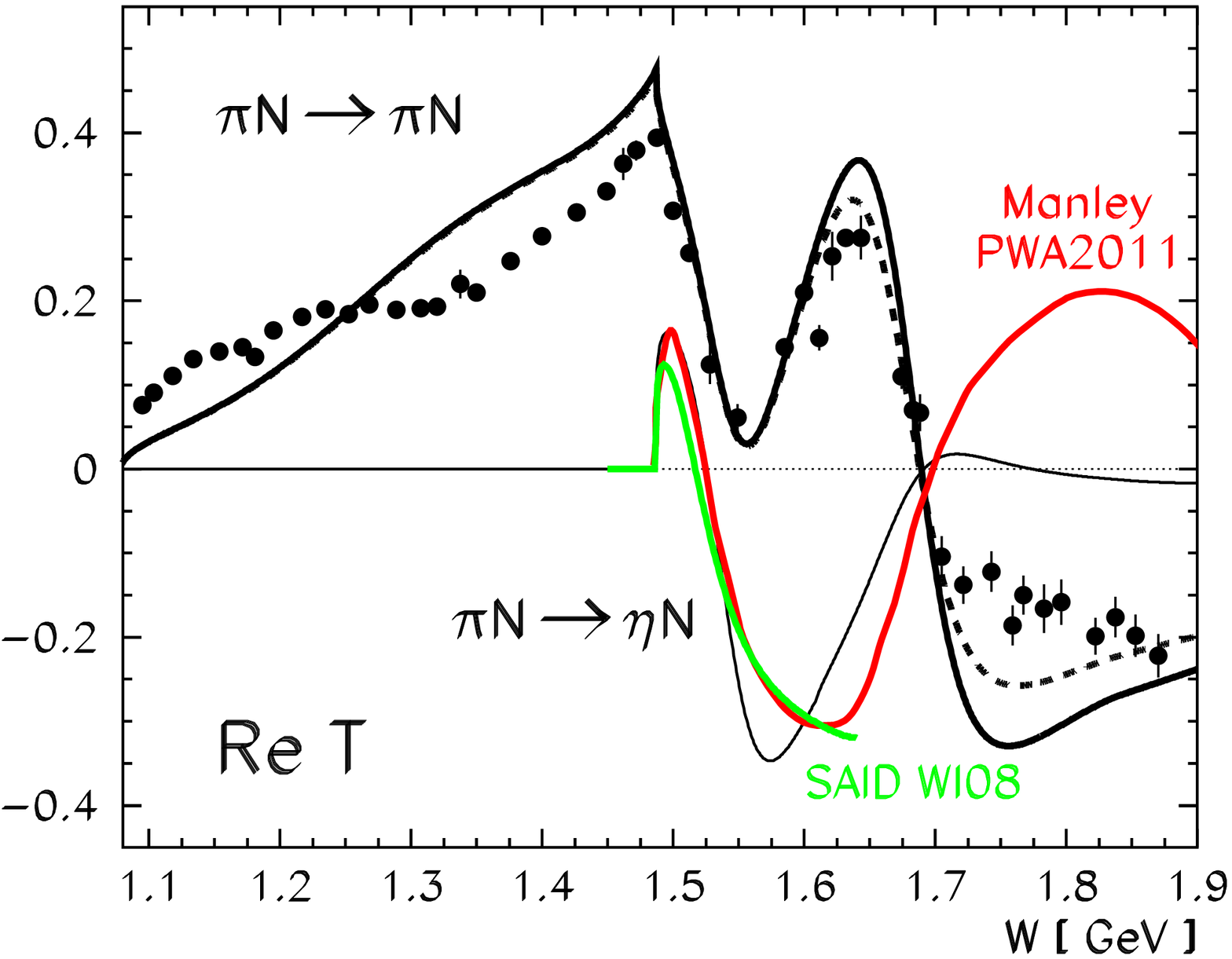}
    \includegraphics[width=0.47\textwidth]{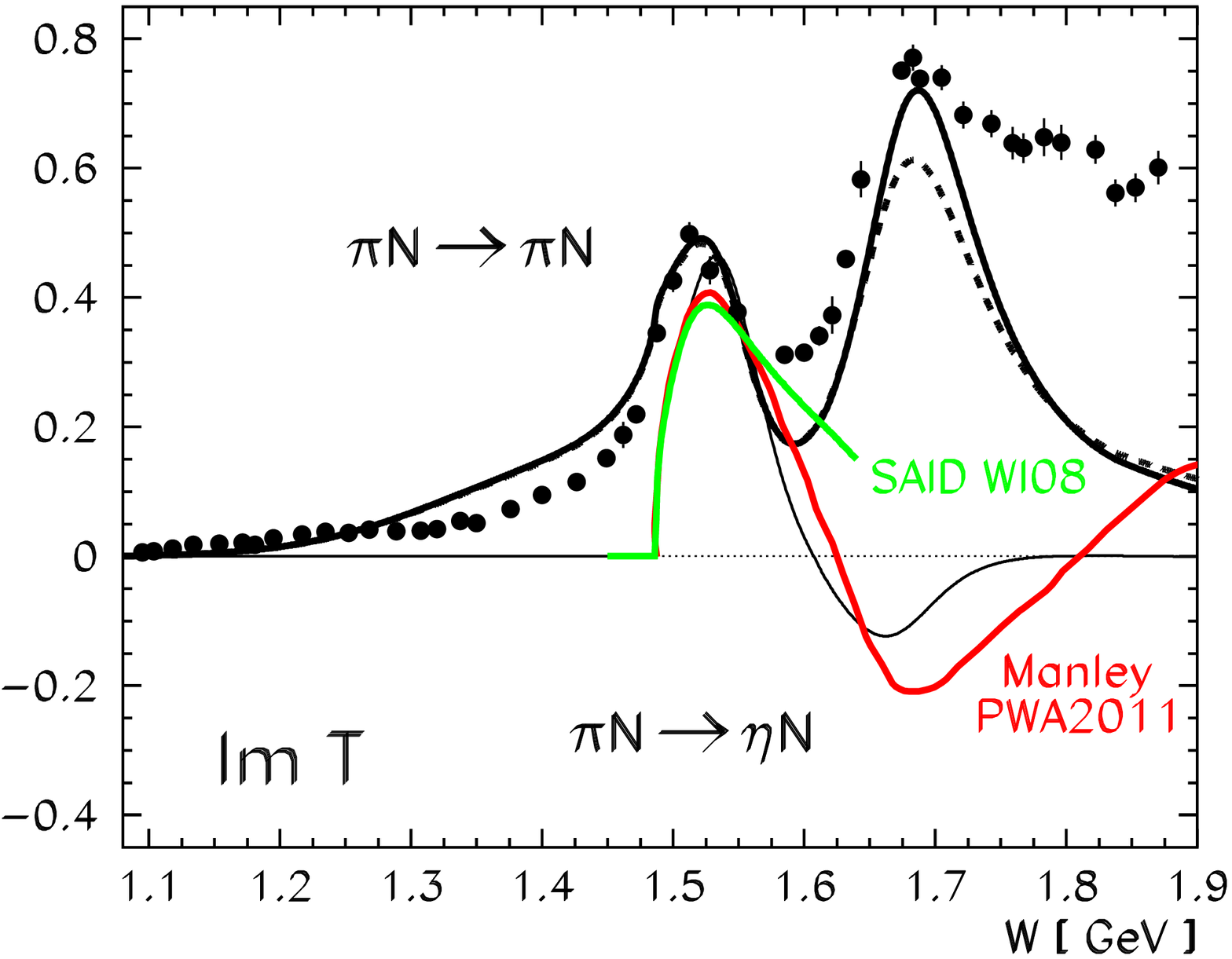}
    \vspace*{-3mm}
    \caption{The real and imaginary part of the scattering
$T$ matrix for the S11 partial wave.  Shown are the elastic
results as well as those for the $\pi N \rightarrow\eta N$ channel.
The solid lines were obtained by using the reduced value of the 
$d$-wave $\pi\Delta$ coupling while the dashed lines correspond to 
the unmodified quark-model values for the baryon-meson couplings.} 
    \label{fig:S11pietaRI}
  \end{center}
\end{figure}

\clearpage

The quark-meson vertices were computed in a SU(3)-extended
Cloudy Bag Model with the quark-meson Lagrangian
$
{\cal L}_{CBM}^{(qM)} 
  = -({\mathrm{i}/ 2{f_M}}) \, \overline{q} \gamma_5 \lambda_a
    q \phi_a \delta\left(r-{R_{bag}}\right)$, $a=1,2,\ldots,8$.
We used $R_{bag} = 0.83\,\mathrm{fm}$, the interaction
parameter $f_\pi = 76\,\mathrm{MeV}$ which reproduces the
experimental $g_{\pi NN}$ coupling, $f_K = 1.2 \, f_\pi$, and
either $f_\eta = 1.2 \, f_\pi$ or $f_\eta = f_\pi$.
The bare masses of $\Phi_{\cal R}$ were adjusted to reproduce
the positions of the resonances.

We have reproduced reasonably well the main features
of the $M_{1-}$ electro-production amplitude from the threshold
up to $W\sim 1700$~MeV (Figs.~\ref{fig:pM1m} and \ref{fig:S11pietaRI}),
as well as the transverse and scalar helicity amplitudes
for $Q^2$ up to $\sim 3$~GeV$^2$/c$^2$. 
Our calculations indicate that the pion cloud plays
an important role, especially in the region of low $Q^2$ (long-range effects).
In the case of the $N^*(1440)$ resonance (Fig.~\ref{fig:hel1520}),
the quark contribution to $A_{1/2}^p$ is positive, while the pion
contribution and the vertex corrections due to meson loops are negative.
These two effects are responsible for the zero crossing 
of the amplitude.  At higher $Q^2$ (short-range physics)
the quark core takes over, rendering $A_{1/2}^p$ positive.

\begin{figure}[htbp]
  \begin{center}
    \includegraphics[width=0.49\textwidth]{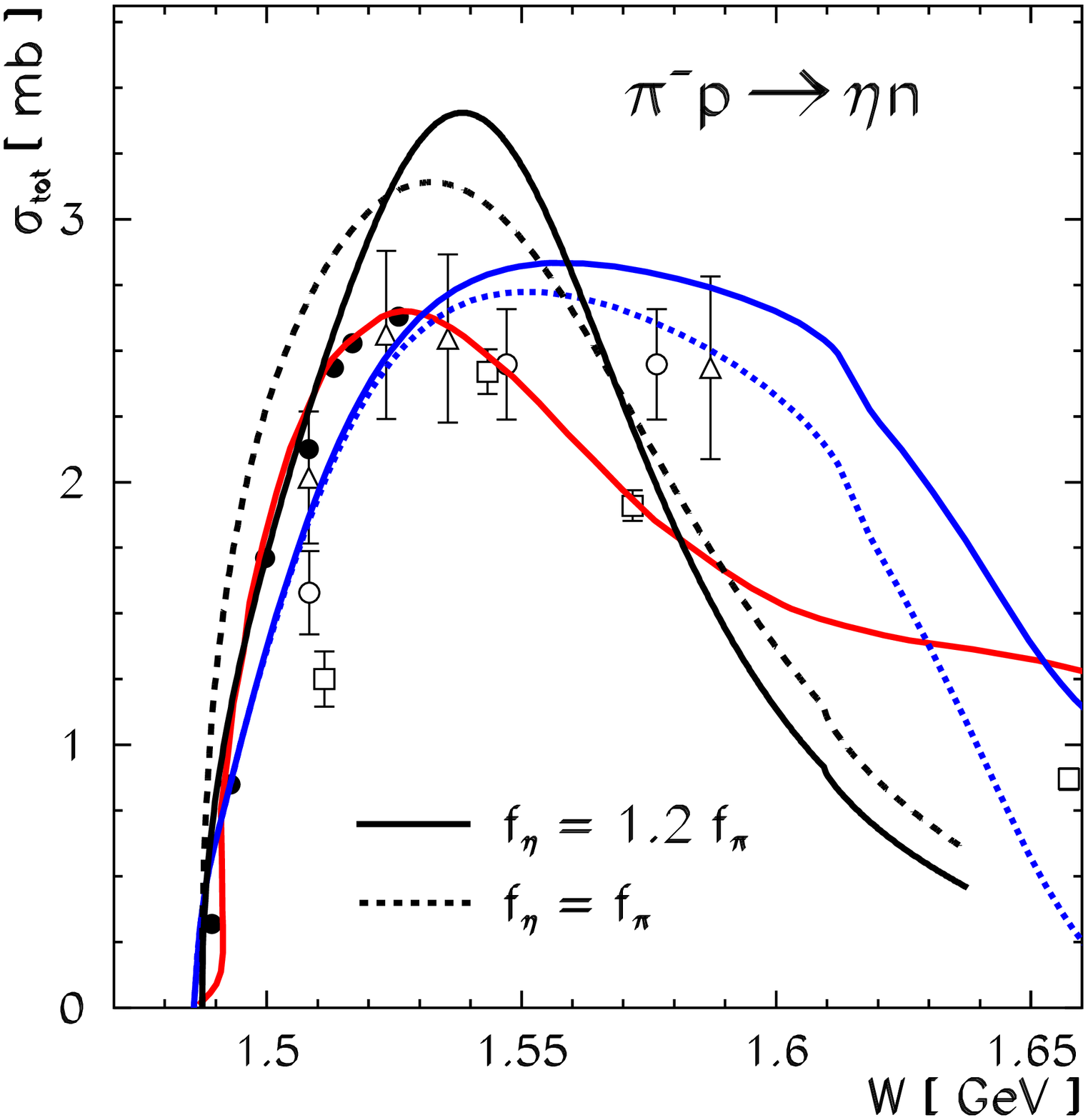}
    \includegraphics[width=0.49\textwidth]{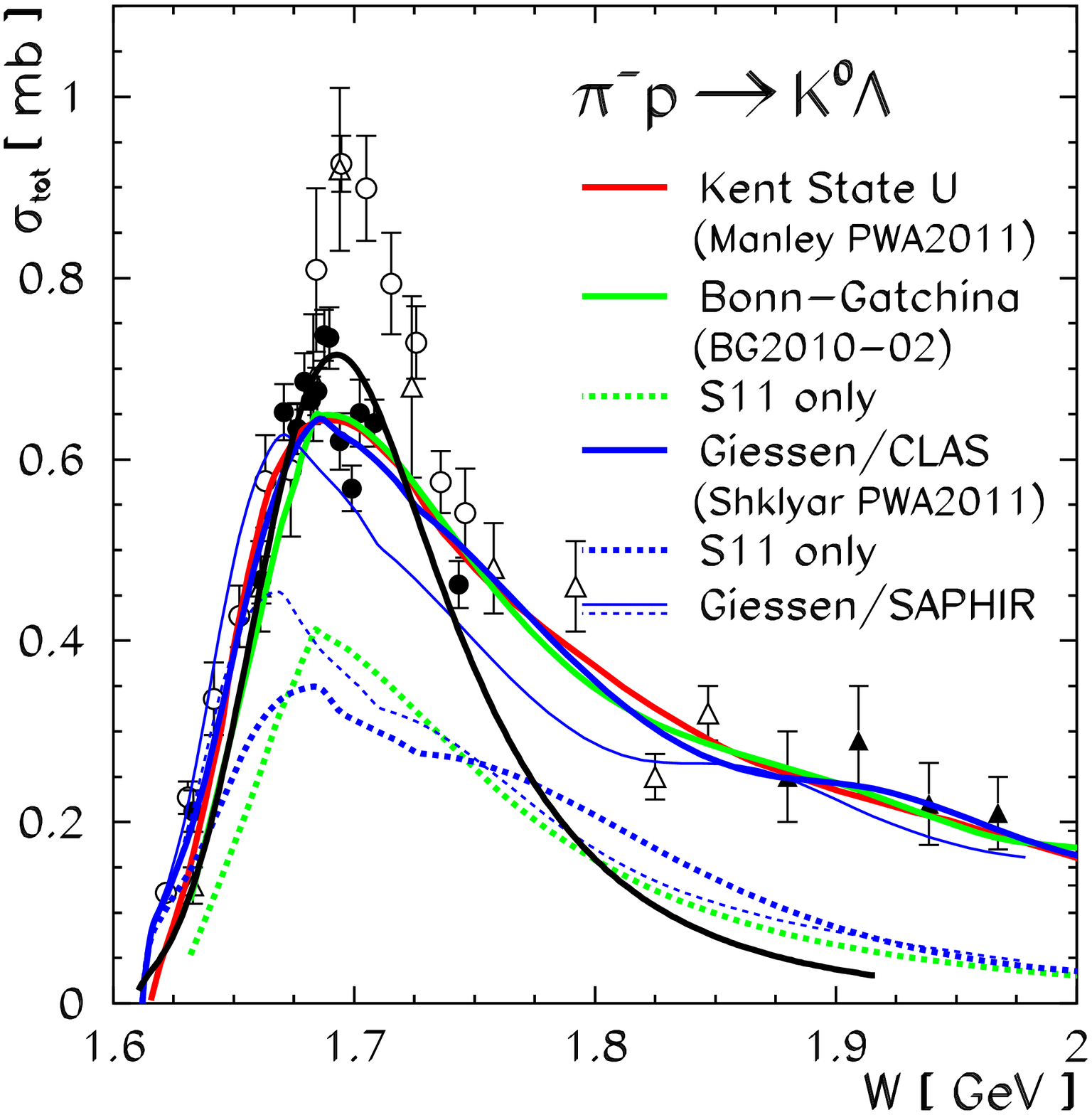}
    \vspace*{-3mm}
    \caption{Left: the S11 contribution to the total 
cross-section for the $\pi^-\,p\rightarrow \eta\, n$ reaction.
Right: total cross-section for the $\pi^-\,p\rightarrow K^0\, \Lambda$
process.  Our result: black solid and dashed lines; other analyses
are color-coded in the figure legend.}
    \label{fig:S11pixs}
  \end{center}
\end{figure}

The results for pion-induced meson production amplitudes
in the S11 partial wave are shown in Fig.~\ref{fig:S11pietaRI}.
Near the $N^*(1535)$ resonance, just above the $\eta$ threshold,
the elastic and inelastic amplitudes are strongly influenced by
the $s$-wave $\eta N$ channel.  Near the $N^*(1650)$ resonance,
additional channels become more relevant,
such as the $\pi\Delta$ with $l=2$, the $K\Lambda$ with $l=0$, 
two channels involving the $\rho$ meson with $l=0$ ($\rho_1N$) 
and $l=2$ ($\rho_3N$), and the $\pi N^*(1440)$ channel with $l=0$.
The S11 contributions to the total cross-section for
the $\pi^-\,p\rightarrow \eta\, n$ and $\pi^-\,p\rightarrow K^0\, \Lambda$
processes are shown in Fig.~\ref{fig:S11pixs} (left and right,
respectively).

\newpage

The results for $\eta$ photoproduction are shown
in Fig.~\ref{fig:E0K} (left).  The behavior is almost
completely driven by the properties of the $N^*(1535)$
resonance and the threshold behavior of the $\eta N$ amplitude,
but is almost insensitive to background processes.
The good overall agreement with the data for $\eta$ production 
strongly supports our conjecture about the dominance of the 
genuine three-quark configuration in the $N^*(1535)$ state.
The absolute value of the $E_{0+}$ amplitude 
in the $K\Lambda$ channel is shown in Fig.~\ref{fig:E0K} (right).
While the cross-section for pion-induced production of $K^+$
appears to be over-estimated in our model, the photo-production
amplitude is smaller than predicted by phenomenological analyses.
This discrepancy remains an open question and represents
a challenge for further investigation.

\begin{figure}[htb]
  \begin{center}
    \includegraphics[width=0.35\textwidth]{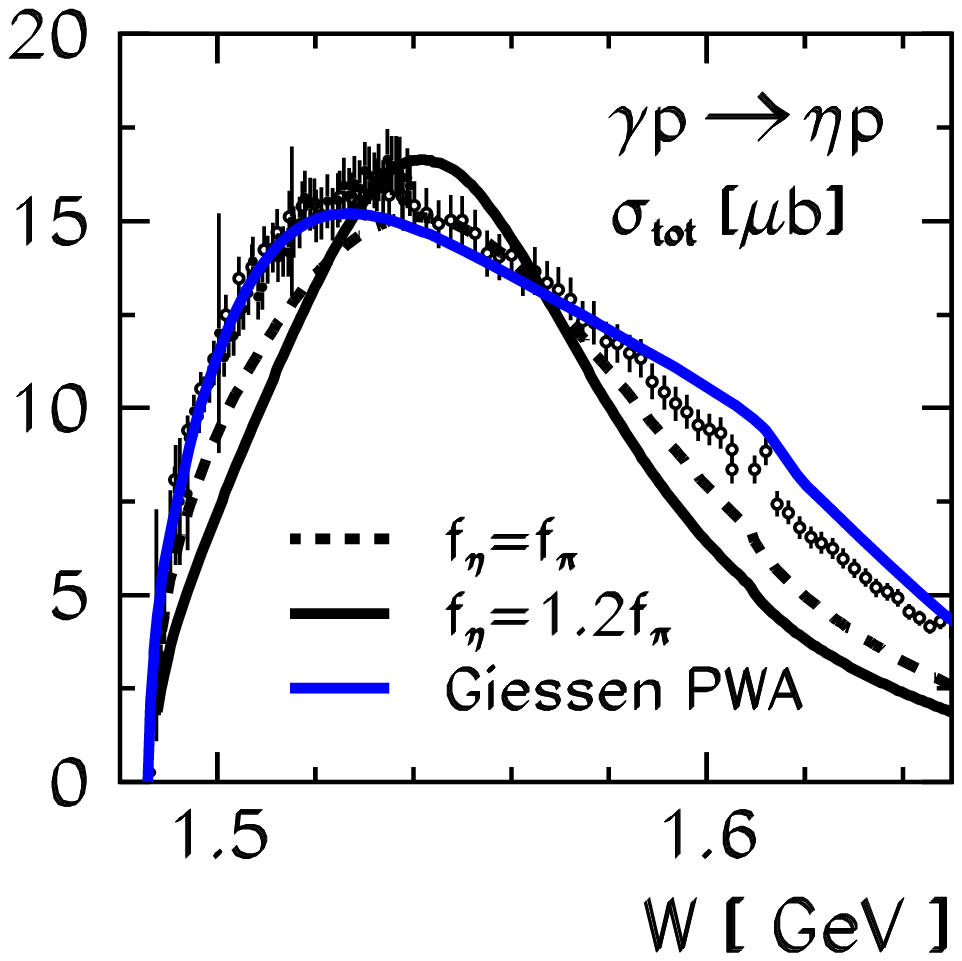}
    \includegraphics[width=0.35\textwidth]{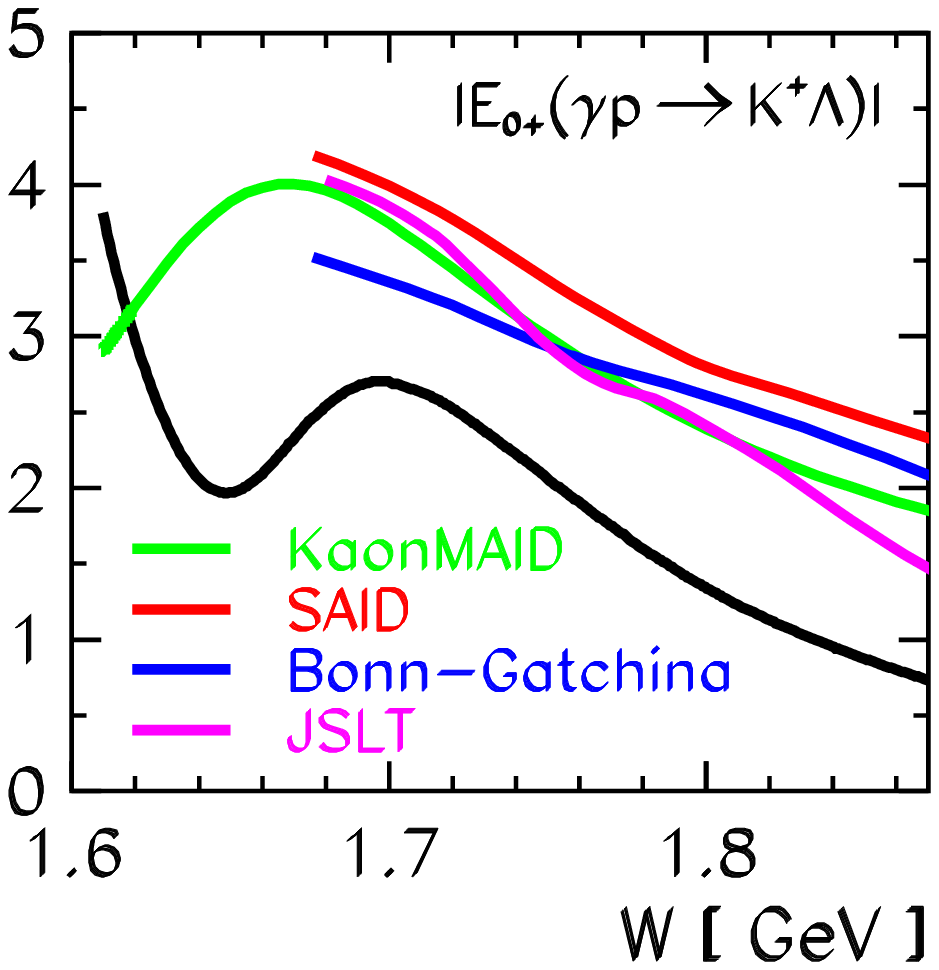}
    \caption{Left: total cross-section for eta photoproduction
on the proton.  Right: the absolute value of the transverse $K^+$
photo-production amplitude $E_{0+}$. Our result: black solid
and dashed lines; other calculations are color-coded in the figure legend.}
    \label{fig:E0K}
  \end{center}
\end{figure}

{\bf Acknowledgement}  The Authors wish to thank
Andrei Sarantsev and Vitaliy Shklyar for providing us
with the most recent data files from their analyses.


%

}  


\end{document}